# Computational investigation of the plasmonic properties of TiN, ZrN, and HfN nanoparticles: The role of particle size, medium, and surface oxidation

Yashar Esfahani Monfared[*] and Mita Dasog[*]

*Department of Chemistry, Dalhousie University, 6274 Coburg Road, Halifax, NS, Canada.*

*\*Email: y.monfared@dal.ca, mita.dasog@dal.ca.*

**Abstract.** Group 4 transition metal nitride (TMN) nanoparticles (NPs) display strong plasmonic responses in the visible and near-infrared regimes, exhibit high melting points and significant chemical stability and thus are potential earth-abundant alternatives to Au and Ag based plasmonic applications. However, a detailed understanding of the relationship between TMN NP properties and plasmonic response is required to maximize their utility. In this study, the localized surface plasmon resonance (LSPR) frequency, bandwidth, and extinction of titanium nitride (TiN), zirconium nitride (ZrN), and hafnium nitride (HfN) NPs were examined as a function of the particle size, surface oxidation, and refractive index of the surrounding medium using finite element method (FEM). A linear redshift in the LSPR frequency and a linear increase in the associated full-width at half maximum (FWHM) was observed with increasing the particle size, oxidation layer thickness, and medium refractive index. We show that the effect of surface oxidation on plasmonic properties of TMN NPs is strongly size-dependent with a significant LSPR redshift, intensity reduction, and broadening in small NPs compared to larger NPs. Furthermore, the performance and efficiency of HfN, ZrN, TiN as well as Au NPs for a narrowband application - photothermal therapy (PTT), and a broadband application - solar energy conversion, was investigated in detail. The results indicate that narrowband and broadband photothermal performance of NPs strongly depend on the particle size, surface properties and in case of narrowband absorption, excitation wavelength.

## 1. INTRODUCTION

Plasmonic nanoparticles (NPs) allow the manipulation of light and energy at the nanoscale and have been applied in surface-enhanced Raman spectroscopy (SERS),[1] label-free biosensing,[2] solar energy conversion,[3] water desalination[4] and photothermal therapy.[5] Photons incident on a metal NP surrounded by a dielectric medium can induce coherent collective oscillation of the free electrons in the NP known as surface plasmons (SPs).[6,7] Excitation of SPs is usually accompanied by a substantial absorption and scattering (collectively known as extinction) of light in discrete optical frequency bands known as localized surface plasmon resonances (LSPRs).[6,7] Extinction characteristics of plasmonic NPs depend on the intrinsic optical characteristics of the material, the NP size, shape, surface chemistry, and the refractive index of the surrounding dielectric medium.[6-8] The performance of plasmonic NPs in many applications, especially absorption-based cases such as photothermal therapy and solar-driven water evaporation is proportional to the amount of light absorbed. Au and Ag are typically used as plasmonic materials due to their strong absorption in the visible and near-infrared (near-IR) wavelength regimes.[9,10] However, Ag NPs are susceptible to surface oxidation and exhibit an overall low surface stability.[9,10] Au NPs exhibit beneficial chemical stability and biocompatibility but succumb to sintering at nanoscale and present a high material cost.[11] Thus, alternatives to these noble metals are being pursued to enable high-temperature and/or commercially scalable applications.[6,10]

Transition metal nitrides,[11] transparent conducting oxides (TCOs),[12] semiconductors,[13] group 13-15 metals,[14] and inter-metallics[15] have been investigated as the potential candidates to replace Au and Ag in many plasmonic applications. Recent experimental investigations have demonstrated the potential of group 4 transition metal nitride (TMN) NPs for plasmonic applications involving both narrowband and broadband light absorption.[16-23] TMN NPs exhibit melting points in excess of 2000 °C and significant chemical inertness.[17,18] Such materials also display high conductivity and corrosion resistance enabling operation in harsh environments.[17, 18]

Previous experimental works have reported varying photothermal efficiency and performance benchmarks for TMN NPs. This variance is hypothesized to be a consequence of different synthetic approaches utilized to generate the nanomaterials, and as a result, variations in size, shape, surface chemistry, polydispersity, and suspension medium of TMN NPs. Computational studies have been performed to understand the plasmonic properties of group 4 TMN NPs,[24-26] yet there is a lack of more in-depth discussions on the plasmonic responses as a function of NP size, shape, surface oxidation, and surrounding medium. Understanding the role of these parameters on the plasmonic response and photothermal properties of TMN NPs may enable better prediction of the performance of particles in different applications, facilitate the characterization of NPs and guide synthetic procedures to target desired plasmonic responses. In this report, we present a computational investigation of the plasmonic response of group 4 TMN NPs (TiN, ZrN, and HfN) including the frequency of the extinction maxima, extinction bandwidths, and extinction efficiency as a function of the NP size, surface chemistry, and surrounding optical medium. The role of such physical characteristics towards photothermal therapy and solar energy absorption was also analyzed.

## 2. COMPUTATIONAL METHOD

We utilized a finite element method (FEM) solver for Maxwell's equations applied with COMSOL Multiphysics to calculate the absorption, scattering, and extinction spectra of TMN NPs. Extinction cross section ($C_{ext}$) were calculated as the summation of absorption and scattering cross sections of NPs due to incident light using equation (1),[27]

$$C_{ext} = \frac{2}{n\sqrt{\varepsilon_0/\mu_0}\,E_{in}^2} \int u_{av}\,dv + \frac{\int |E_{far}|^2 d\Omega}{n\,E_{in}^2} \qquad (1)$$

where $\varepsilon_0$ is permittivity of free space, $\mu_0$ is permeability of free space, $E_{in}$ is the incident electric field, $u_{av}$ is the power absorbed by the NP in the form of ohmic losses, $E_{far}$ is the far field component of the outgoing wave calculated by the Stratton-Chu formula and n is the refractive index of the medium. The first term in equation (1) determines the absorption cross section by integrating the resistive heating over the volume of NP and dividing by the incident power density. The second term determines the scattering cross section using the normalized integral of the far field component of the outgoing electromagnetic energy flux over a boundary surrounding the NP.[27] Normalized extinction of a particle can be defined as extinction cross section ($C_{ext}$) divided by the geometric cross section of NP. Here we report the computational values in terms of normalized extinction or absolute cross sections.

A perfectly matched layers (PML) boundary conditions as well as ultra-fine mesh size was applied to the models. Three-dimensional models (nanospheres) were meshed via the built-in meshing algorithm in COMSOL with a maximum element size of 0.1r (r is the radius of nanospheres) for simulating NPs. In addition, rigorous convergence analysis was done for each simulation to evaluate the accuracy of the results. To consider the effect of surface oxidation, NPs were simulated as nitride-core oxide-shell 3D structures. To simulate the material optical properties, real and imaginary part of TiN, ZrN, HfN, $TiO_2$, $ZrO_2$ and $HfO_2$ dielectric constants ($\varepsilon = \varepsilon_r + i\varepsilon_i$) as a function of excitation wavelength were obtained from litereature.[26,28-30] The real ($\varepsilon_r$) and imaginary ($\varepsilon_i$) part of the dielectric function of TMN NPs between 300 nm and 800 nm is shown in Figure 1. Data analyses and fittings were done using Matlab R2019a.

The computational model has been verified by comparing our simulation results to previous theoretical calculations[26] and validated by our previous experimental observations[31,32] in group 4 TMN NPs. The results of the verification studies are reported in Figure S1 in the Supplementary Information. Our computational model has been verified by comparing the absorption and scattering spectra of TiN, ZrN and HfN NPs with previously reported spectra of TMN NPs with 100 nm diameter size.[26] As our calculated spectra agree well with previously reported data derived from Mie theory (and DFT calculations) for group 4 TMN NPs suspended in water[26], we assume our methodology remains valid for other particle dimensions, surrounding media, and surface characteristics.

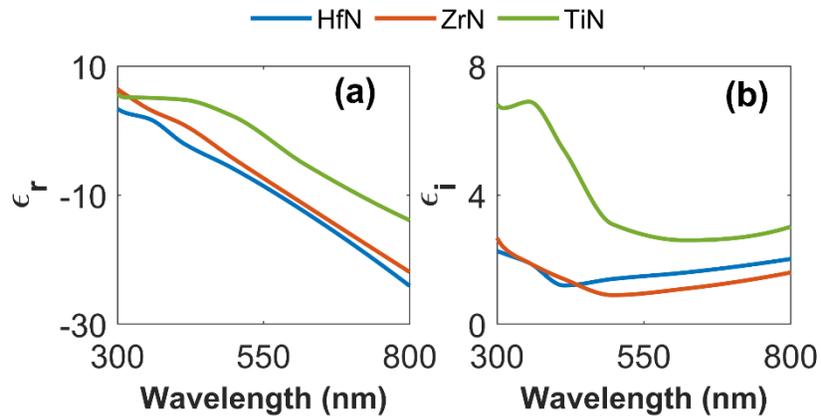

**FIGURE 1.** (a) Real and (b) imaginary components of the dielectric function of the group 4 TMNs between 300 and 800 nm.

## 3. PLASMONIC RESPONSE OF TMN NPS

### 3.1. Effect of the NP size

The effect of the NP size on the frequency of the LSPR and the associated spectral full-width at half maximum (FWHM), and also the ratio between scattering and absorption was calculated for NPs with a diameter ranging between 10 and 50 nm. This range was chosen as many of the experimental reports yield TMN NPs in this size regime. The maxima redshifted with increasing particle size for all TMN NPs investigated: HfN (Figure 2a), ZrN (Figure 2b) and TiN (Figure 2c). The shift was linear with the particle size growth as shown in Figure 2d ($R^2$ values for the linear fits were determined to be 0.996, 0.997 and 0.989 for HfN, ZrN and TiN NPs, respectively). The magnitude of the LSPR redshift with increasing particle size was observed to be the highest for HfN where the peak shifted from 440 to 465 nm as the size increased from 10 to 50 nm, respectively. The resonance shifted from 484 to 500 nm for ZrN and from 585 to 600 nm for TiN NPs as the size increased from 10 to 50 nm. The absorption and scattering cross sections of HfN, ZrN, and TiN NPs between 10 – 50 nm is reported in Supplementary Information (Figure S2).

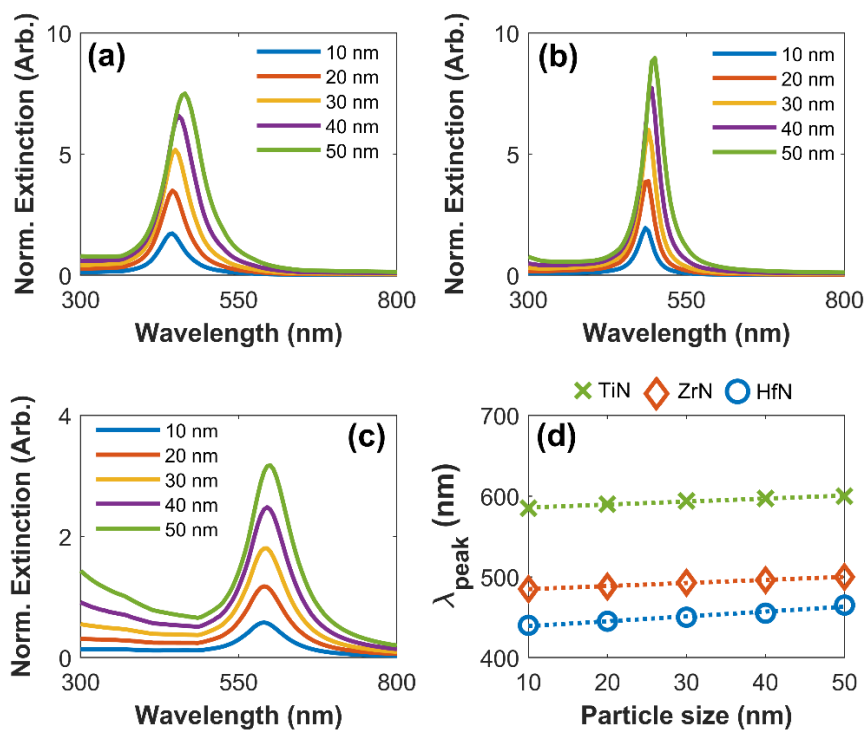

**FIGURE 2.** Normalized extinction of (a) HfN, (b) ZrN and (c) TiN NPs of varying size suspended in water as a function of wavelength. (d) LSPR maxima wavelength of TMN NPs as a function of the particle size. Note that all of the data points in (d) are fitted with linear fit functions and $R^2$ values of linear fits for HfN, ZrN and TiN are 0.996, 0.997 and 0.989, respectively.

The effect of the NP size on the broadness of the LSPR (FWHM) and the ratio of scattering to absorption was studied and is presented in Figure 3. TiN exhibited the broadest LSPR (ranging between 82 – 92 nm) relative to ZrN (28 – 38 nm) and HfN (49 – 68 nm) for NP size between 10 and 50 nm (Figure 3a). For 10 nm particles, the FWHM of TiN (82 nm) was observed to be almost three times larger than that of ZrN (28 nm). The data indicates that FWHM increases with particle size (Figure 3a), however, this effect was observed to be the highest for HfN (19 nm broadening of FWHM) compared to 10 nm broadening for ZrN and TiN.

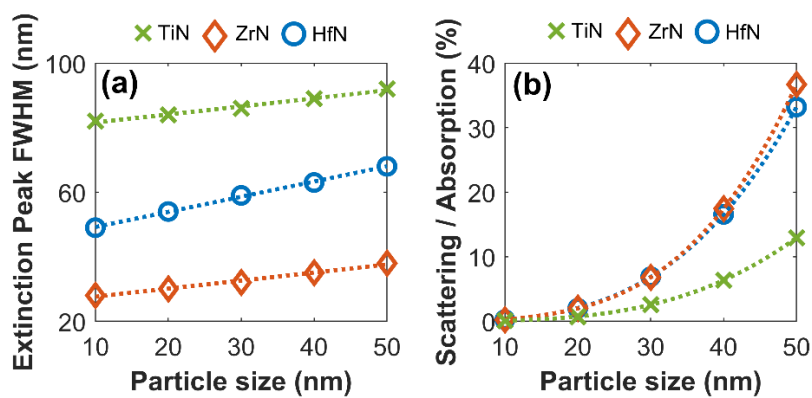

**FIGURE 3.** (a) LSPR FWHM for TMN NPs as a function of particle size. (b) Scattering to absorption ratio as a function particle size for TMN NPs. Note that the data points are fitted with linear fit function in (a) and cubic function in (b). $R^2$ values of linear fits for HfN, ZrN, and TiN are 0.997, 0.991 and 0.986, respectively.

The ratio between scattering and absorption can significantly increases by an increase in the particle size (Figure 3b). To determine this relationship, the ratio between maximum scattering cross section and maximum absorption cross section of NPs was calculated from Figure S2 in the Supplementary Information. As the particle size increases from 10 to 50 nm, scattering to absorption ratio increases from ~0.2% (for all three particles) to 33.2, 36.6, and 12.9% for HfN, ZrN, and TiN NPs, respectively. This trend demonstrates that for smaller particles absorption is the dominant extinction mechanism and the scattering contribution to extinction is negligible. The data also suggest that ZrN has the highest scattering to absorption ratio for large NPs between group 4 TMNs, a trend which is consistent with previous theoretical predictions for 100 nm transition metal NPs.[26]

## 3.2 Effect of the surrounding medium

The correlative frequency shift of the observed plasmon resonance of plasmonic NPs as a function of the refractive index of the surrounding medium is a phenomenon commonly utilized to provide evidence for the plasmonic-origin of an optical signature.[16,18] Any variations in the refractive index of the medium close to the surface of NPs will lead to a change in surface charges and in turn, the intensity and position of the extinction peak.[33,34] We examined the extinction cross section spectra of TMN NPs as the refractive index of the surrounding medium was varied between 1.30 to 1.45 (Figure S3), values consistent with typical solvents used to suspend plasmonic TMN NPs in experimental reports. Increasing the medium index from 1.30 to 1.45, the LSPR of 30 nm TMN NPs were observed to redshift and exhibited small increases in magnitude and FWHM (Figure 4). As the size of particles does not change the dynamics of extinction in different mediums, we only reported the results for 30 nm TMN NPs. The variation of LSPR frequency was observed to be linear with increase in the medium index for HfN, ZrN and TiN, a manner consistent with previous experimental observations.[16,18] It should be noted that HfN showed the largest variations (23 nm) with a fixed change in medium RI from 1.3 to 1.45, compared to TiN (17 nm) and ZrN (13 nm). The previous experimental study performed by our group showed largest shift in LSPR frequency in HfN NPs when dispersed in medium with varying refractive index consistent with the calculations.[16] TiN NPs were found to have the smallest shift in the experimental studies.[16] The discrepancy between experimental and calculated data could be due to difference in particle size or surface oxidation. The sensitivity of HfN NPs to medium RI demonstrate the potential of HfN for RI sensing applications. The of medium RI on broadness of the LSPR was also studied (Figure 4b). The LSPR FWHM exhibited by TMN NPs increases linearly when the medium RI increases from 1.30 to 1.45. The magnitude of broadening was higher in TiN and HfN NPs with 9 nm and 8 nm increase in FWHM, respectively, compared to 3 nm increase for ZrN NPs.

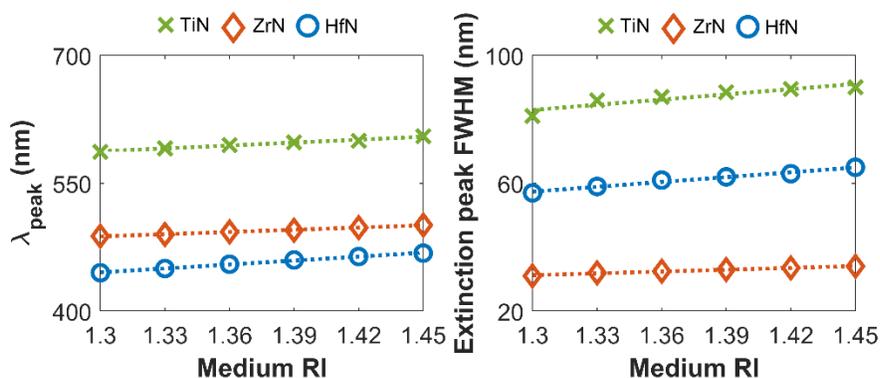

**FIGURE 4.** (a) LSPR wavelength of TMN NPs with a 30 nm size as a function medium refractive index. Note that all of the data points in (a) are fitted with linear fit functions and $R^2$ values of the linear fits for HfN, ZrN and TiN are 0.996, 0.998 and 0.993, respectively. (b) LSPR FWHM for TMN NPs with 30 nm size as a function of medium RI. Note that all the data points in (b) are fitted with linear fit functions and $R^2$ values of the linear fits for HfN, ZrN and TiN are 0.998, 0.995 and 0.976, respectively.

### 3.3. Effect of the surface oxidation

Surface oxidation or formation of a self-passivating layer on the NP surface is frequently observed in the synthesis of TMN NPs.[16-18] Previously, it was shown that surface oxidation can cause a significant reduction in resonant energy and consequently, a redshift in absorption/scattering maxima of ZrN NPs.[18] Here, we focused our analysis on the effect of surface oxidation on the plasmonic properties of TiN, ZrN, and HfN NPs. NPs were simulated as being constructed with a metal nitride-core (with diameter of $d$) and oxide-shell (with thickness of $D$) (Figure 5a) and were considered to be suspended in deionized water. The LSPR modes were observed to redshift with increasing the thickness of the oxidation layer. Increasing oxide layer thickness resulted in an attenuation of the LSPR mode for all three TMN NPs (Figure S4). The data suggests that the smaller NPs display heightened sensitivity to surface oxidation in their plasmonic response (Figure 5b, 5d, and 5f). For example, for a smaller TiN NP (nitride core, $d = 6$ nm), the LSPR shifted from 580 nm to 710 nm (130 nm shift) as the oxide shell thickness was increased from 0 to 3 nm whereas the LSPR for a large TiN particle ($d = 30$ nm) was observed to shift from 590 nm to 620 nm (only 30 nm shift).

Similarly, the FWHM evolution was larger for smaller NPs with increasing oxidation when compared to the larger ones (Figure 5c, 5e and 5g). The FWHM of smaller (nitride core, $d = 6$ nm) TiN NPs increased from 80 nm to 112 nm while FWHM for larger NPs ($d = 30$ nm) only increased from 86 to 90 nm. Furthermore, it can be inferred from normalized extinction spectra of TMN NPs that the magnitude of the LSPR decreases with an increase in oxidation layer thickness (Figure S4), a trend which is consistent with previous experimental and theoretical observations.[18] This reduction in extinction is dependent on the nitride core size and is significantly stronger for smaller TMN NPs. Similar trends were observed for HfN and ZrN NPs but the rates of the LSPR redshift, increase in FWHM and reduction of the resonance magnitude were significantly different. For example, for a small HfN NP (6 nm) extinction peak shifts from 455 nm 515 nm (60 nm peak variation) as the oxide shell thickness was increased from 0 to 3 nm while the extinction peak wavelength for a large HfN particle (30 nm) shifted from 465 nm to 480 nm (only 15 nm variation). The results suggest that TiN NPs LSPR frequency and FWHM are more sensitive to surface oxidation compared to HfN and ZrN NPs.

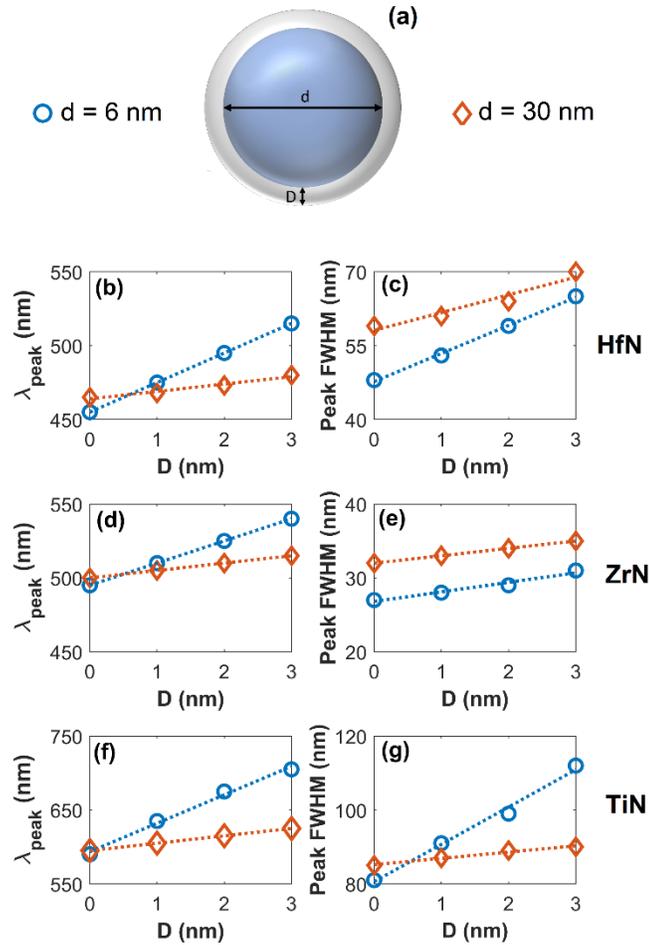

**FIGURE 5.** (a) Schematic of the oxidation geometry in the simulations, particles are simulated in a three-dimensional nitride-core (with a diameter of *d*) and oxide-shell (with a thickness of *D*) geometry. (b, d and f) wavelength of the LSPR as a function of oxide layer thickness for HfN, ZrN and TiN NPs, respectively. (c, e and g) FWHM of extinction peak as a function of oxide layer thickness for HfN, ZrN and TiN NPs, respectively. Note that all of the data points in (b)-(g) are fitted with linear fit functions and $R^2$ values of the linear fits for are between 0.998 and 0.978. The blue circles illustrate the data points for core diameter of 6 nm and orange diamonds illustrate the data points for core diameter of 30 nm.

# 4. PHOTOTHERMAL APPLICATIONS

## 4.1. Narrowband absorption

Strong manipulation of light via absorption and scattering processes make plasmonic materials useful in many medical applications including drug delivery, biosensing, bioimaging, and cancer therapy as well as non-medical applications such as optical energy conversion and nanocatalysis.[5,35-37] Here, we examine non-radiative localized photothermal applications and in particular two main applications of TMN NPs which are based on narrowband absorption for photothermal therapy and broadband absorption application for solar energy conversion. These applications utilize plasmonic materials to transform incident optical energy into localized heat with high efficiency.[35,36] Generally the performance and efficiency of such conversion is governed by the absolute amount of light absorption (heat generation) and the ratio between light absorption and light scattering (the efficiency of the energy conversion). The efficiency of a NP in terms of absorbing the light energy is usually defined as the ratio of absorption cross section and extinction cross section of the particle within the FWHM of the light source. The actual heat generation is a function of absorption cross section of NP within FWHM of the light source. The efficiency of TMN NPs for narrowband applications is strongly depend on the excitation wavelength and bandwidth of the source (e.g. light-emitting diode or laser). Performance is dictated by the overlap between the NP extinction spectra and the source emission spectra.

Photothermal therapy (PTT) relies on accumulation of plasmonic NP on or inside the tumors and then they are irradiated to localize heat at the tumor location, resulting in selective hyperthermia and irreversible damage to the tumor.[5] Near infrared (NIR) light is usually used in different biomedical and PTT applications as they can penetrate deep into the body due to the reduced absorption by biological tissues. In most of the PTT studies, researchers utilized light sources in the first biological transparency window which is located in the 700 to 1000 nm wavelength range known as the traditional first NIR window (NIR-I window).[5,35] Recently, PTT with light sources in the second NIR window (1000–1400 nm wavelength, NIR-II window) have emerged for biomedical applications.[5] Use of the NIR-II can be preferable versus use of the visible or NIR-I windows as light absorption is even lower for biological samples within the NIR-II window which can reduce the risk of damage to the healthy cells during therapy.[5] While visible light (wavelength < 700 nm) cannot penetrate deep into the skin, it has been shown to be effective in skin cancer treatments.[35] Here, we study narrowband absorption performance of TMN NPs for four excitation wavelengths which are usually used in PTT studies. We examine the efficiency and performance of TiN, ZrN, and HfN NPs at 532 nm (visible-PTT),[35] 635 nm (visible-PTT),[5] 808 nm (NIR-I PTT),[35] and 1064 nm (NIR-II PTT)[5].

To study the heat generation capabilities and efficiencies of TMN NPs for PTT applications, we defined the photothermal light absorbance (PTT$_{abs}$) and photothermal light-to-heat conversion (photothermal absorption to photothermal extinction) efficiency (PTT$_{eff}$) as,

$$\text{PTT}_{\text{abs}} = \int_{\lambda_{PTT} - 2.5\,nm}^{\lambda_{PTT} + 2.5\,nm} C_{abs}(\lambda)/\sigma \; d\lambda \qquad (2)$$

$$\text{PTT}_{\text{eff}}\,(\%) = \frac{PTT_{abs}}{PTT_{ext}} \times 100 \qquad (3)$$

where, $\lambda_{PTT}$ is photothermal excitation wavelength, $C_{abs}(\lambda)$ is wavelength-dependent absorption cross section, $\sigma$ is geometrical cross section of NP and $PTT_{ext}$ is the photothermal light extinction defined as,

$$PTT_{ext} = \int_{\lambda_{PTT} - 2.5\,nm}^{\lambda_{PTT} + 2.5\,nm} C_{ext}(\lambda)/\sigma \; d\lambda \qquad (4)$$

$\lambda_{PTT}$ or PTT laser excitation wavelengths of 532, 635, 808, or 1064 nm were explored in our studies. We considered a case wherein all NPs were 30 nm in size, suspended in aqueous medium, and compared the performance and efficiency of these NPs with similar size Au NPs. In our calculations, we considered the possible effect of source FWHM by performing the integral over a 5 nm spectral range. Photothermal absorbance of Au and TMN NPs are compared at different excitation wavelengths (Figure 6).

For 30 nm particle size, Au NPs showed the highest absorption at 532 nm followed by ZrN as the LSPR of these particles (528 nm and 495 nm for Au and ZrN, respectively) was close to the excitation wavelength (Figure 6a). TiN displayed the highest absorption at 635, 808, and 1064 nm (Figure 6b, c, and d). The data indicates that TiN NPs can absorb and generate almost 10 times as much heat than Au NPs, and simultaneously demonstrate a superior efficiency for optical to thermal conversion close to its resonance wavelength (635 nm). HfN and ZrN were also observed to absorb better than Au NPs at this size and wavelengths between 635 – 1064 nm. Similar simulations were conducted with smaller NPs (6 nm in size), and different performances and trends were observed (Figure S5). For example, HfN can absorb more light than TiN and ZrN at 808 nm and 1064 nm (over NIR) which is consistent with recent experimental observations.[31] The data thus suggest that photothermal performance of particles are strongly depend on the particle size and excitation wavelength.

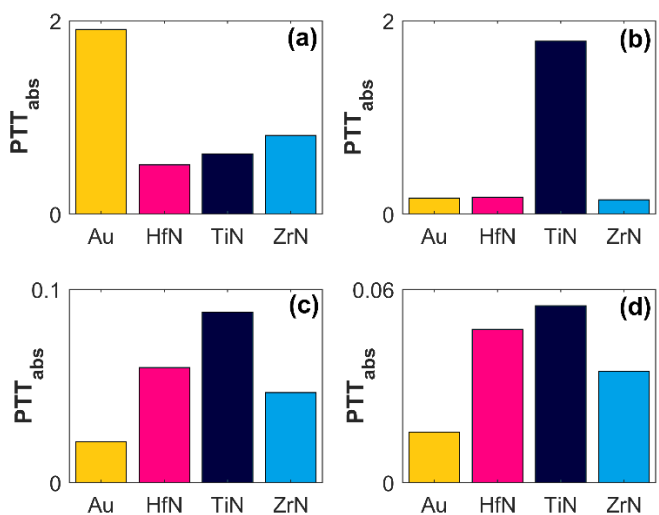

**FIGURE 6.** Comparison of light absorption of TMN and Au NPs with 30 nm size at (a) 532 nm, (b) 635 nm, (c) 808 nm and (d) 1064 nm.

To further understand the role of the TMN NP size on light absorption and PTT light-to-heat conversion efficiency, a more detailed investigation was performed on TiN NPs as a function of size at excitation wavelengths of 635 and 808 nm (Figure 7). An increase in NP size from 10 nm to 50 nm resulted in more light absorption but a decrease in the absorption to extinction ratio (light-to-heat conversion efficiency) at both operation wavelengths. This light-to-heat conversion efficiency decrease is due to the increase in scattering cross section as the NPs increase in size, similar to trend observed in Figure 3b. Thus, the NPs absorb more light but also scatter more light which in turn contribute to a lower conversion efficiency for larger NPs.

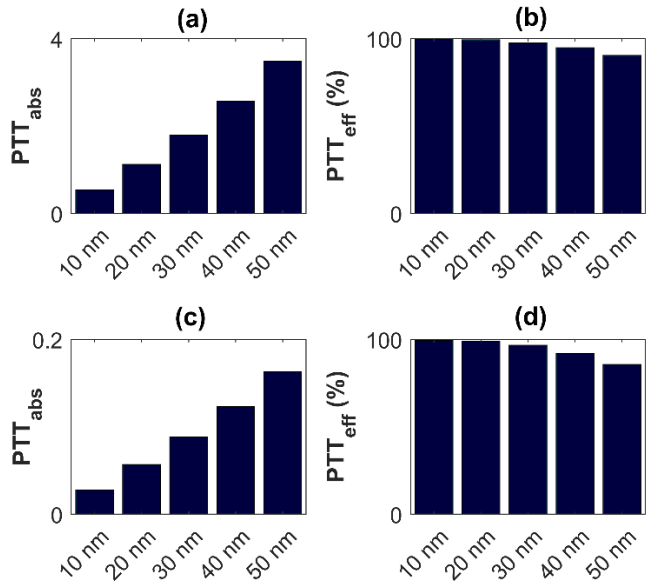

**FIGURE 7.** Comparison of (a) photothermal light absorbance and (b) photothermal light-to-heat conversion efficiency of TiN with different sized NPs at 635 nm. Comparison of (c) photothermal light absorbance and (d) photothermal light-to-heat conversion efficiency of TiN with different sized NPs at 808 nm.

## 4.2. Broadband absorption

Plasmonic NPs can be used to absorb solar insolation which can be used for solar energy conversion, water purification, and desalination applications.[4,32,38] For wavelengths less than 280 nm, solar radiation is blocked by the ozone layer and for wavelengths longer than 1500 nm, the infrared radiation of solar spectrum is almost completely absorbed in 1 mm layer of water.[38] Therefore, the effective bandwidth of solar spectrum for any NP suspended in water is between 280 nm and 1500 nm. To compare the performance and efficiency of TMN NPs in absorbing solar energy and converting it to heat, we define a broadband solar absorption ($S_{abs}$) and solar light-to-heat conversion (solar absorption to solar extinction) efficiency ($S_{eff}$) for any NP suspended in water as,

$$S_{abs} = \int_{280\,nm}^{1500\,nm} \frac{c_{abs}(\lambda)}{\sigma} \times S(\lambda)\ d\lambda \qquad\qquad (5)$$

$$S_{eff}\ (\%) = \frac{S_{abs}}{S_{ext}} \times 100 \qquad\qquad\qquad (6)$$

where $S(\lambda)$ is the normalized solar spectrum and $S_{ext}$ is the broadband solar extinction defined as,

$$S_{ext} = \int_{280\,nm}^{1500\,nm} \frac{c_{ext}(\lambda)}{\sigma} \times S(\lambda)\ d\lambda \qquad\qquad (7)$$

We first consider a case where all particles are 30 nm in size and suspended in water. The normalized absorption cross section of HfN, ZrN, TiN, and Au NPs as well as the normalized solar radiation spectrum is depicted in Figure 8a for reference. HfN has the highest broadband solar absorption (Figure 8b) which means it can absorb and convert more solar energy than the other materials at this size, a trend which is consistent with recent experimental findings.[32] The solar light-to-heat conversion efficiency of TMN and Au NPs are almost similar at this size, ranging from 94% to 97% (Figure 8c) which means most of the solar energy will be absorbed by NPs.

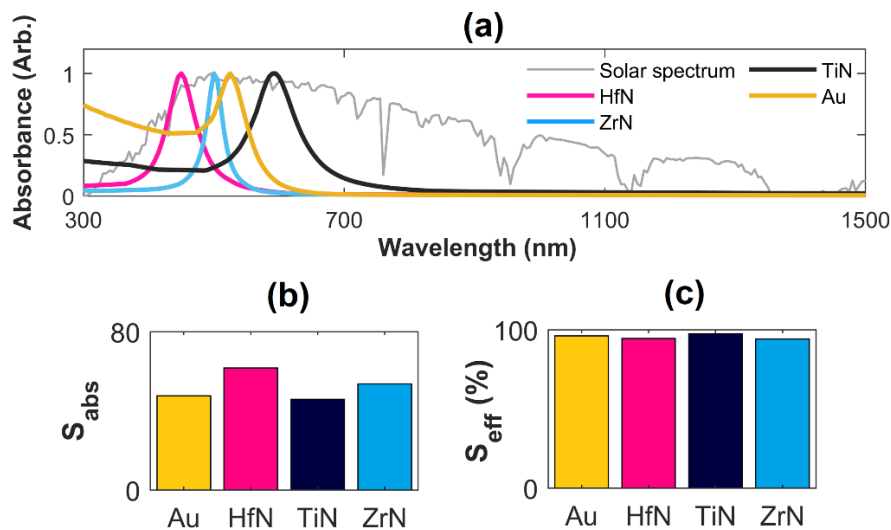

**FIGURE 8.** (a) The normalized absorbance of HfN, ZrN, TiN and Au NPs with 30 nm particle size as well as the normalized solar radiation spectrum from 280 nm to 1500 nm. (b) Comparison of broadband solar absorption and (c) solar light-to-heat conversion efficiency of TMN NPs with 30 nm size

We also compared the broadband solar absorption of TMN and Au NPs where the size of particles is 50 nm in Figure S6. HfN broadband solar absorption increased by 52% by increasing the particle size from 30 nm to 50 nm while TiN broadband solar absorption increased by 74%. This means TiN solar absorption enhances more than HfN NPs with a fixed increase in particle size. The large growth rate of TiN solar absorption suggest that TiN may perform better than HfN in broadband absorption applications for larger particles (>50 nm). Moreover, the solar light-to-heat conversion efficiency of TiN is significantly better than HfN and Au for large particles (Figure S5b). Therefore, while HfN performs far better as small to medium size particle (10 – 30 nm) in broadband solar absorption, as the particles become larger (>50 nm) TiN starts to become a better candidate for broadband applications. The effect of the size on broadband solar absorption and solar light-to-heat conversion efficiency of HfN NPs are studied in more detail, and similar trend to narrowband application is observed as shown in Figure S7. By increasing the size of NPs, the actual amount of solar absorption increases significantly (Figure S6a) while solar light-to-heat conversion decreases (Figure S6b). This trend means a significant portion of solar energy will be scattered by the large HfN NPs.

Finally, we explored the role of surface oxidation on photothermal heat generation of TMN NPs in broadband solar energy absorption applications. Here we repeat the computations for TiN, ZrN and HfN NPs where a 2 nm oxide shell is present on the surface of the particles. We chose small particles (6 nm) for this study as the effect of surface oxidation is more significant in smaller NPs, as discussed in section 3.3. This shell thickness is representative of that observed in experimental works with TMN NPs.[16,18] Note that the solar light-to-heat conversion efficiency at this size is close to 99% for all three TMN NPs which means almost all of the solar energy will be absorbed by NPs without any significant scattering. Presence of an oxide shell reduces the broadband solar absorption for all TMN NPs compared to when no oxide is present (Figure 9).

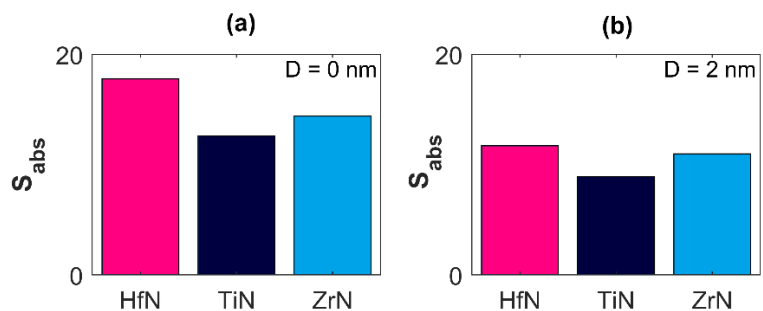

**FIGURE 9.** Comparison of broadband solar absorption of TMN NPs with a nitride core size of 6 nm, (a) with no oxide shell and with (b) 2 nm oxide shell.

However, the reduction in solar absorption is different for different TMN NPs. For example, HfN broadband solar absorption reduced by 34% by increasing the oxide shell thickness from 0 to 2 nm while TiN broadband solar absorption reduced by 29%. This means HfN solar absorbance reduces more than TiN NPs with a fixed change in oxide shell thickness growth. The lower rate of reduction in TiN solar absorbance is possibly due to the fact that TiN LSPR broadens significantly with surface oxidation compared to other TMN NPs, as demonstrated in section 3.3. These results clearly indicate that performance of TMN NPs in both narrowband and broadband photothermal applications are strongly dependent on their size and surface characteristics.

## 5. CONCLUSIONS

The effects of the particle size, extent of surface oxidation, and medium refractive index on the LSPR frequency and FWHM of plasmonic group 4 TMN NPs was investigated in detail. We observed a linear redshift of the LSPR and increase in associated FWHM by increasing the particle size, oxidation layer thickness and medium refractive index. The calculations also suggested that smaller NPs are far more sensitive to surface oxidation. For a small TiN particles (6 nm nitride core diameter), the LSPR shifted from 580 nm to 710 nm (almost 130 nm) by increasing the oxidation thickness from 0 to 3 nm, while the LSPR for large TiN particle (30 nm) shifted from 590 nm to only 620 nm (about 30 nm). Furthermore, the efficiency and performance of TMN and Au NPs were compared for narrowband and broadband photothermal-based applications including photothermal therapy (PTT) and solar energy conversion. The results indicate that smaller HfN NPs (~10 nm) and medium to larger TiN NPs (30 – 50 nm) are potential candidates for PTT applications with a performance and an efficiency better than Au NPs in near infrared. In the case of solar energy conversion, HfN exhibited superior performance relative to other TMN and Au NPs for small to medium size particles (10 – 30 nm) while TiN is a potential candidate in case of large particles (>50 nm).

**SUPPLEMENTARY MATERIAL**

See supplementary material for Figures S1 – S7.

**ACKNOWLEDGMENTS**

The authors acknowledge funding from the Natural Science and Engineering Research Council of Canada (NSERC Discovery Grant # 2017-05143) and New Frontiers in Research Fund (NFRFE-2018-00051). CMC Microsystems Canada and Dalhousie University are thanked for providing the access to Comsol Multiphysics and Matlab softwares, respectively.

# Supplementary Information

# Computational investigation of the plasmonic properties of TiN, ZrN, and HfN nanoparticles: The role of particle size, medium, and surface oxidation

Yashar Esfahani Monfared[*] and Mita Dasog[*]

*Department of Chemistry, Dalhousie University, 6274 Coburg Road, Halifax, NS, Canada.*

*\*Email: y.monfared@dal.ca, mita.dasog@dal.ca.*

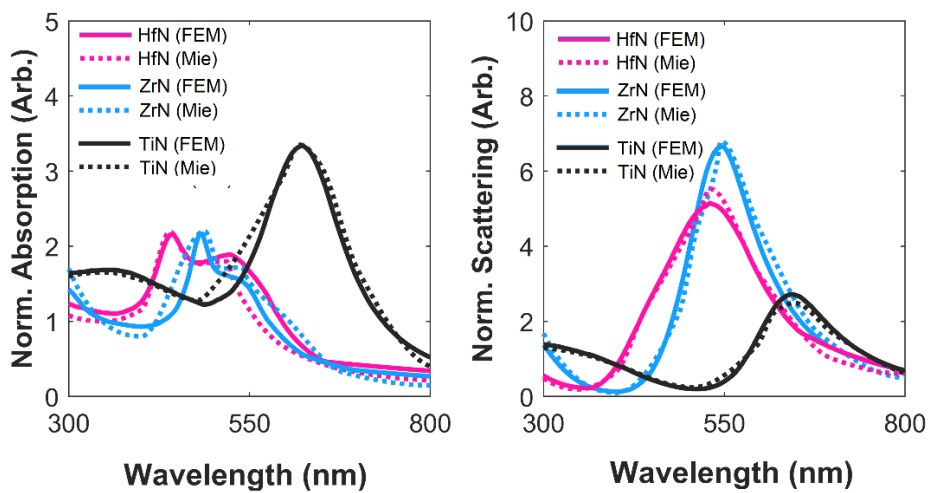

**Figure S1**. Comparison between normalized absorption spectra (left panel) and normalized scattering spectra (right panel) of HfN, ZrN and TiN NPs with 100 nm diameter size suspended in DI water, computed with Mie theory and finite element method. Note that the solid lines represent absorption/scattering spectra obtained using finite element method simulations in our paper, and dotted plots represent absorption/scattering spectra based on Mie theory calculations obtained by Kumar *et al*. Reference: M. Kumar, N. Umezawa, S. Ishii, T. Nagao, *ACS Photonics* **3**, 1, 43-50 (2016)

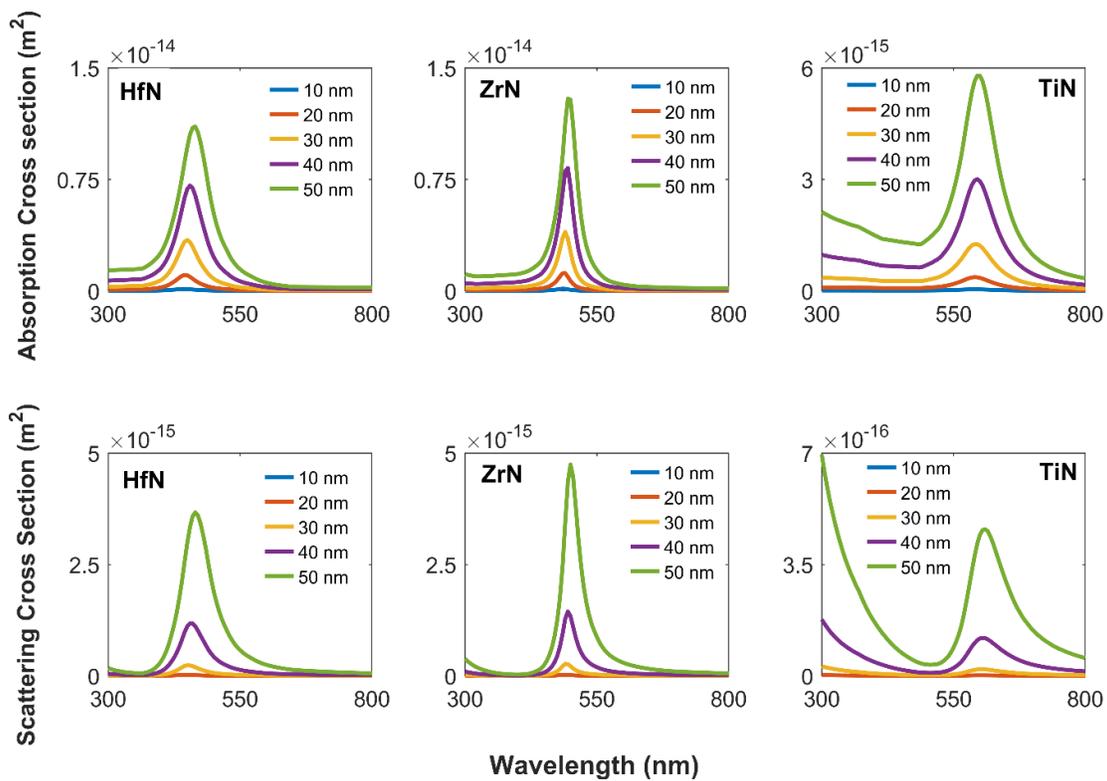

**Figure S2**. Absorption cross sections (top panel) and scattering cross sections (bottom panel) of HfN, ZrN and TiN NPs as a function of wavelength for different particle size (10 nm to 50 nm). Note that the particles are considered to be suspended in DI water in all simulations.

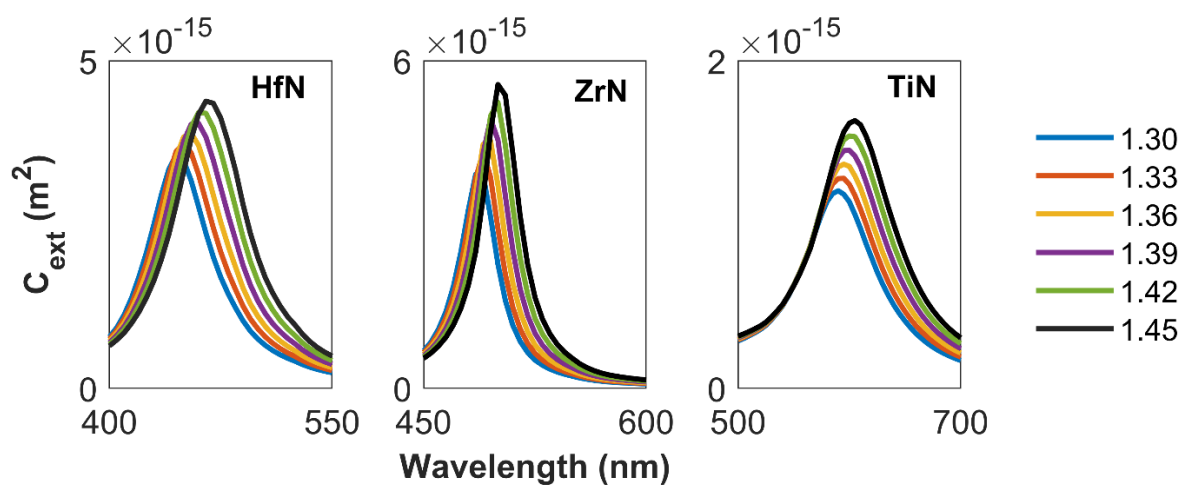

**Figure S3**. Extinction cross section of HfN (left panel), ZrN (middle panel), and TiN (right panel) NPs suspended in different media as a function of wavelength for 30 nm nanospheres.

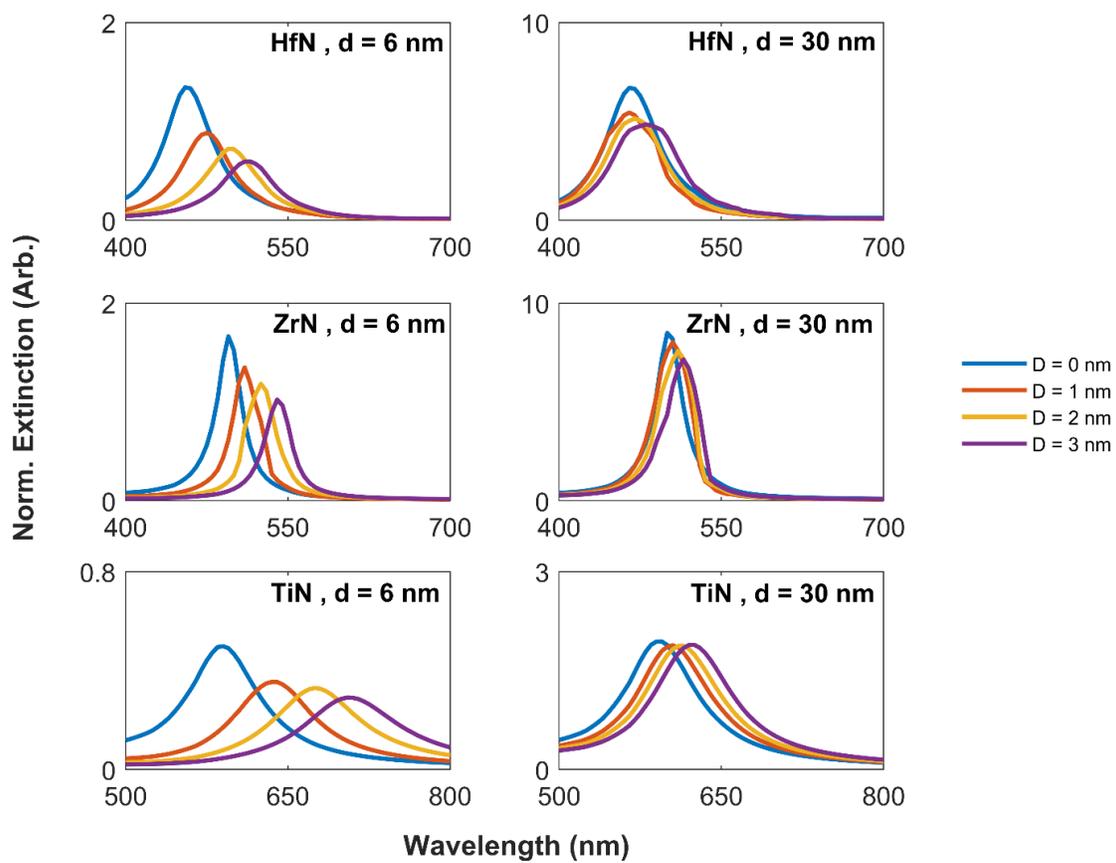

**Figure S4**. Normalized extinction of TMN nanoparticles with a core size of 6 nm (left panels) and 30 nm (right panels) as a function of wavelength for different oxide-layer thicknesses (D) values.

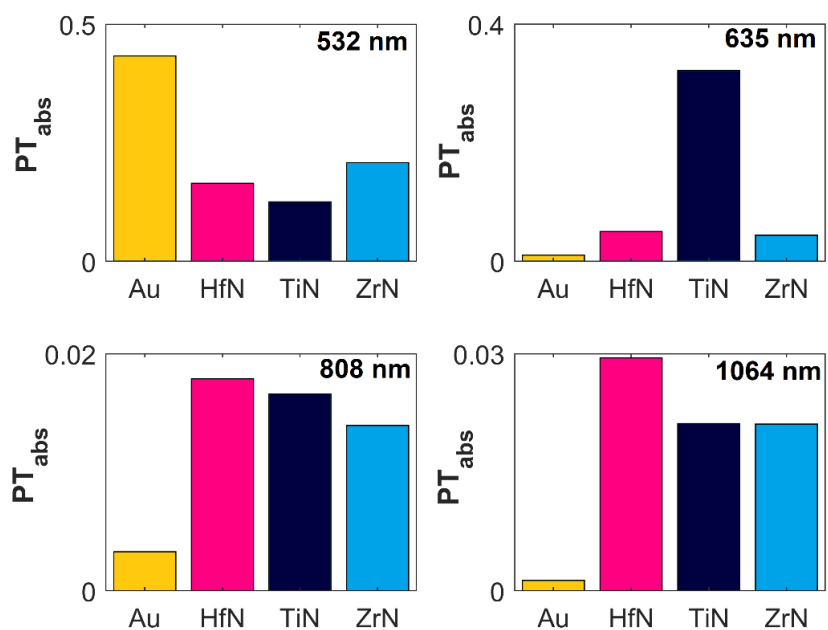

**Figure S5**. Comparison of photothermal absorption of TMN and Au NPs with 6 nm size at the excitation wavelengths of 532, 635, 808, and 1064 nm.

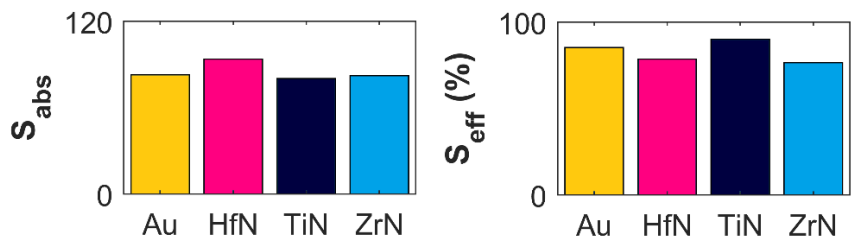

**Figure S6.** Comparison of broadband solar absorption (left panel) and solar light-to-heat conversion efficiency (right panel) of TMN and Au NPs with 50 nm size.

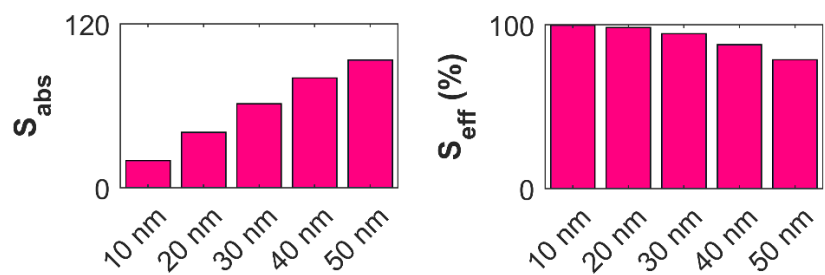

**Figure S7.** Comparison of broadband solar absorption (left panel) and solar light-to-heat conversion efficiency (right panel) of HfN NPs with different particle size.